\documentclass[a4paper]{jpconf}
\usepackage{graphicx}

\usepackage{amssymb}
\usepackage{amsmath}
\usepackage{graphicx}
\usepackage{dcolumn}
\usepackage{bm}

\usepackage{citesort}

\begin{document}
\title{Predictions for Neutrinoless Double-Beta Decay in the 3+1 Sterile Neutrino Scenario}

\author{C. Giunti$^{1}$, E. M. Zavanin$^{1,2,3}$}

\address{
$^1$
INFN, Sezione di Torino, Via P. Giuria 1, I--10125 Torino, Italy
\\
$^2$
Department of Physics, University of Torino, Via P. Giuria 1, I--10125 Torino, Italy
\\
$^3$
Instituto de F\'\i sica Gleb Wataghin, Universidade Estadual de Campinas - UNICAMP,
\\
Rua S\'ergio Buarque de Holanda, 777, 13083-859 Campinas SP Brazil
}

\ead{giunti@to.infn.it, zavanin@gmail.com}

\begin{abstract}
In this proceeding we present predictions of the
effective Majorana mass $|m_{\beta\beta}|$
in neutrinoless double-$\beta$ decay
for the standard $3\nu$ mixing case
and for the 3+1 neutrino mixing case
indicated by the short-baseline anomalies.
We have taken into account the uncertainties of the
neutrino mixing parameters determined by oscillation experiments.
We obtained that the predictions for $|m_{\beta\beta}|$
in the cases of $3\nu$ and 3+1 mixing are quite different,
in agreement with previous discussions in the literature,
and that
future measurements of
neutrinoless double-$\beta$ decay
and
the total mass of the three lightest neutrinos in cosmological experiments
may distinguish the $3\nu$ and 3+1 cases
if the mass ordering is determined by oscillation experiments.

\end{abstract}

\section{Introduction}
\label{sec:intro}

Are neutrinos Dirac or Majorana particles?
This question cannot be probed in neutrino oscillation experiments.
Since the lepton number is conserved in oscillation experiments there is no difference between a Dirac and a Majorana neutrino. However, the Majorana
nature of neutrinos can be investigated in experiments in which there is violation of the lepton number. The most promising type of experiment
in which it can be probed are the neutrinoless double-beta
decay experiments, in which the total lepton number is violated by two units (see the recent review in Ref.~\cite{Bilenky:2014uka}).


Using the current measurements of neutrino squared-mass differences and mixing angles
it is possible to predict the possible range of values for the
effective Majorana mass
$|m_{\beta\beta}|$
in neutrinoless double-beta decay for different orderings of the neutrinos masses.


The plan of this proceeding is to present some of the results obtained in Ref. \cite{Giunti:2015kza}.
We  will discuss in Section \ref{sec:3nu} the predictions for
$|m_{\beta\beta}|$
in the standard $3\nu$ framework,
taking into account the two possible normal and inverted mass orderings and in Section \ref{sec:3p1} we will comment how these predictions are modified
in the 3+1 mixing framework. We will also discuss the possibility of distinguishing the $3\nu$ and the 3+1 mixing framework analyzing the predictions of the effective
Majorana mass in comparison with the sum of the three lightest neutrino masses, that can be obtained by cosmological measurements.

\section{Three-Neutrino Mixing}
\label{sec:3nu}

In the standard three-neutrino ($3\nu$) mixing framework,
the effective Majorana mass in neutrinoless double-beta decay
is defined as
\begin{equation}
|m_{\beta\beta}| = \left| m_{1} |U_{e1}|^2  +  m_{2} |U_{e2}|^2 e^{i\alpha_2} +  m_{3} |U_{e3}|^2 e^{i\alpha_3} \right|.
\label{mbb3nu}
\end{equation}
The elements $U_{ek}$ ($k=1,2,3$) of the mixing matrix,
which quantify the mixing of the electron neutrino with the three massive neutrinos,
can have unknown complex phases,
which generate the two complex phases
$\alpha_2$
and
$\alpha_3$
in Eq.~(\ref{mbb3nu}).

In this work we use the results of the global fit
of solar, atmospheric and long-baseline reactor and accelerator
neutrino oscillation data presented in
Ref.~\cite{Capozzi:2013csa} considering the two possible orderings for the neutrinos masses. The case in which
the lightest neutrino mass is associated with the first mass eigenstate ($m_{min}$ = $m_{1}$) is called Normal Ordering (NO) and the
case in which the lightest neutrino mass is associated with the third mass eigenstate ($m_{min}$ = $m_{3}$) is called Inverted Ordering (IO).

\begin{figure}[t]
\begin{center}
\begin{minipage}{14pc}
\includegraphics[width=14pc]{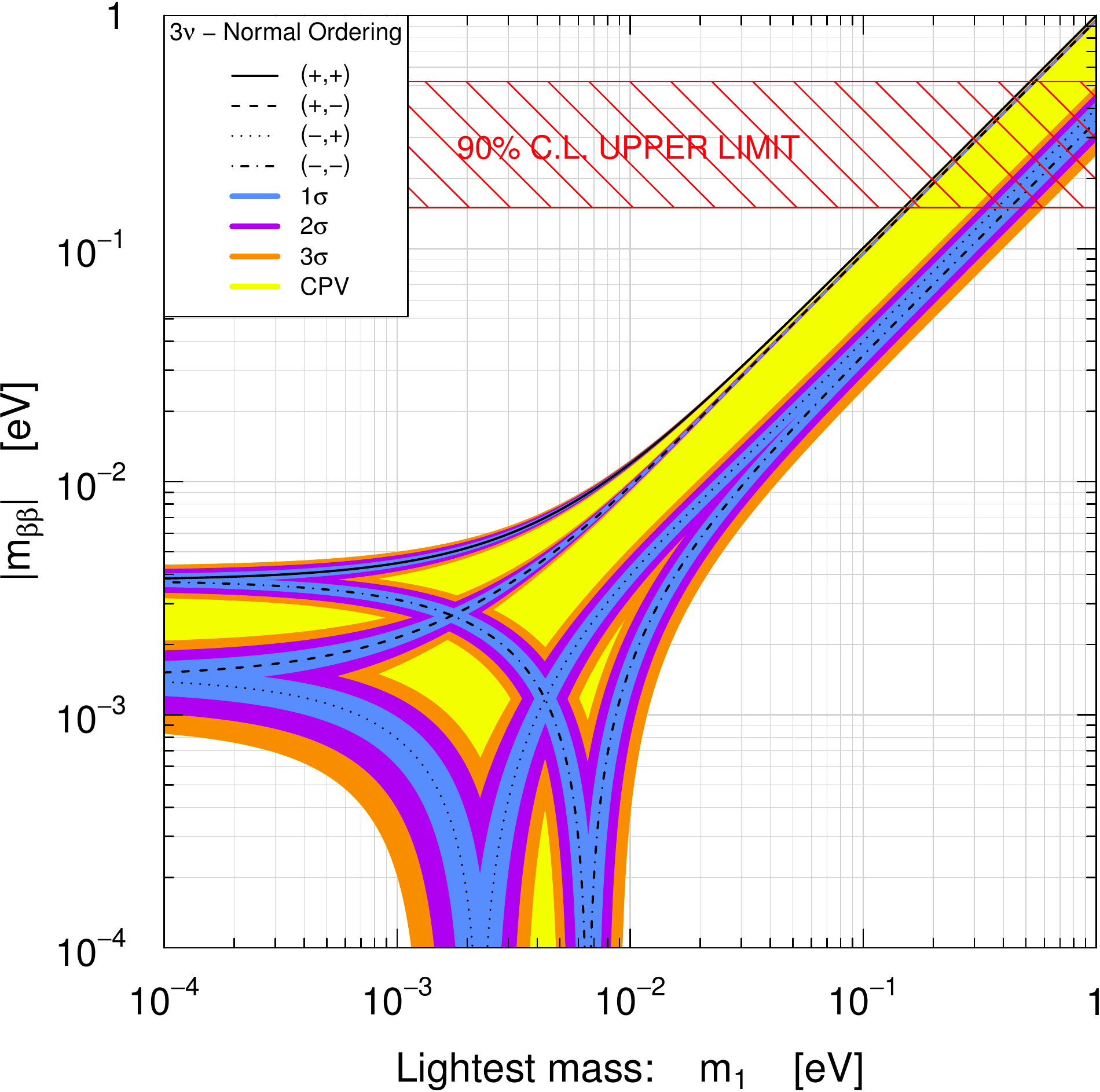}
\end{minipage}\hspace{2pc}%
\begin{minipage}{14pc}
\includegraphics[width=14pc]{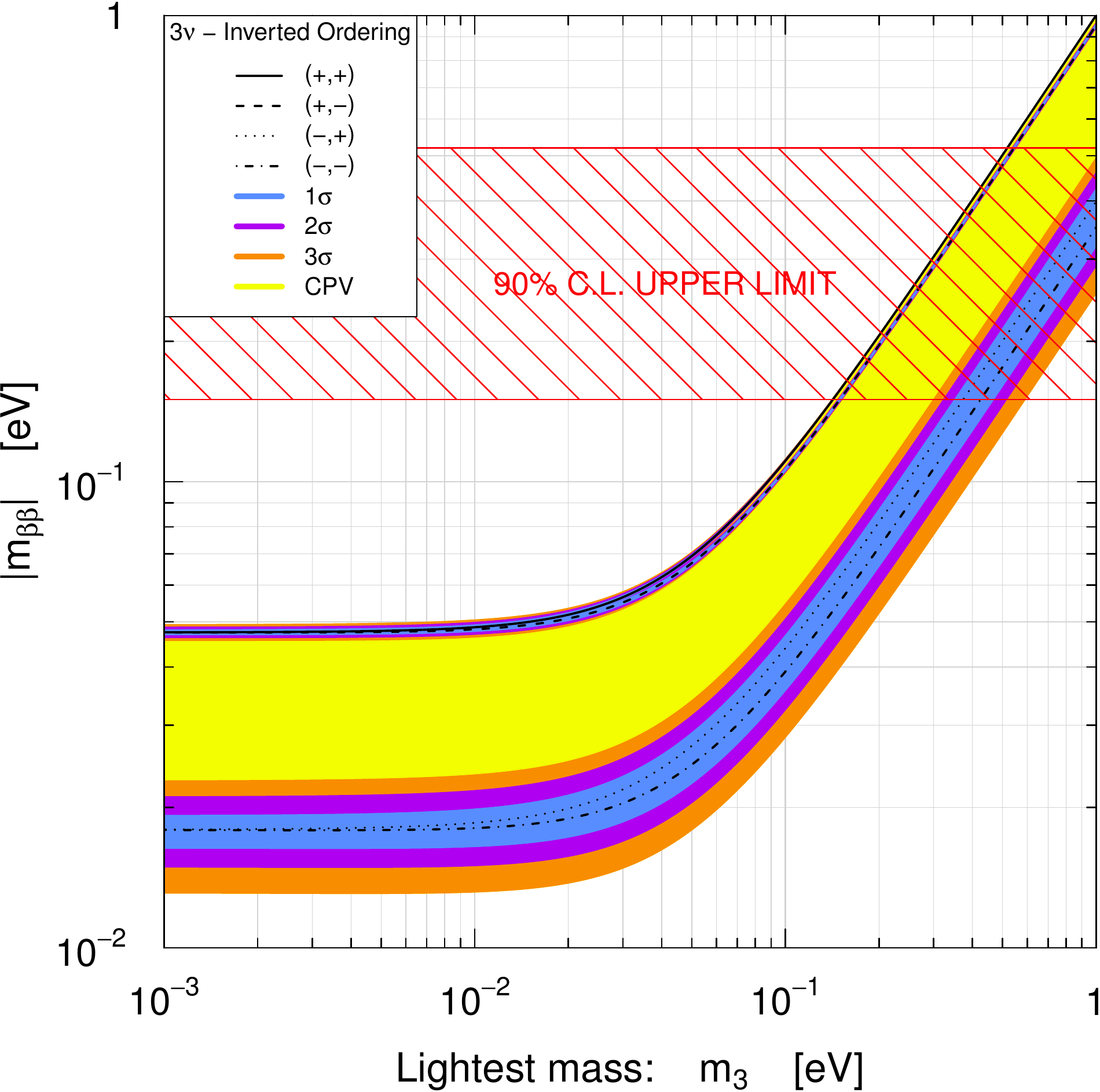}
\end{minipage}
\caption{Left part: Value of the effective Majorana mass $|m_{\beta\beta}|$
as a function of the lightest neutrino mass $m_{1}$
in the case of $3\nu$ mixing with Normal Ordering. Right part: Value of the effective Majorana mass $|m_{\beta\beta}|$
as a function of the lightest neutrino mass $m_{3}$
in the case of $3\nu$ mixing with Inverted Ordering.
The signs in the legend indicate the signs of
$e^{i\alpha_{2}}, e^{i\alpha_{3}} = \pm1$
for the four possible cases in which CP is conserved.
The intermediate yellow region is allowed only in the case of CP violation.
The 90\% upper limit is explained in the main text.}
\label{3numbbvsmin}
\end{center}
\end{figure}

\begin{table}[b!]
\caption{Ranges of
$m_{1}$ for which there can be a complete cancellation of the
three partial mass contributions to $|m_{\beta\beta}|$
for the best-fit values (b.f.) of the oscillation parameters
and at
$1\sigma$,
$2\sigma$ and
$3\sigma$
in the case of $3\nu$ mixing with Normal Ordering.
}
\centering
\renewcommand{\arraystretch}{1.2}
\begin{tabular}{ccccc}
\hline
\hline
&
b.f.
&
$1\sigma$
&
$2\sigma$
&
$3\sigma$
\\
\hline
$m_{1}\,[10^{-3}\,\text{eV}]$
&
$2.3 - 6.6$
&
$1.9 - 7.2$
&
$1.6 - 8.0$
&
$1.3 - 9.0$
\\
\hline
\hline
\end{tabular}
\label{tab:3nu-nor}
\end{table}

The left part of Figure~\ref{3numbbvsmin}
shows the best-fit values
and the
$1\sigma$,
$2\sigma$ and
$3\sigma$ allowed intervals
of the effective Majorana mass $|m_{\beta\beta}|$ in Eq.~(\ref{mbb3nu})
as functions of
the lightest mass $m_{1}$ for the NO scheme, where we have plotted separately the
allowed bands for the four possible cases in which CP is conserved
($\alpha_{2}, \alpha_{3} = 0, \pi$).
The areas between the CP-conserving curves
correspond to values of
$|m_{\beta\beta}|$ which are allowed only in the case of CP violation
\cite{Bilenky:2001rz,Pascoli:2002qm,Joniec:2004mx,Pascoli:2005zb,Simkovic:2012hq,Pascoli:2013fiz}.
We calculated the confidence intervals using the $\chi^2$ function
\begin{equation}
\chi^2_{3\nu} = \chi^2(\Delta{m}_{\text{SOL}}^2) +\chi^2(\Delta{m}_{\text{ATM}}^2) + \chi^2(\sin^2\vartheta_{12})+ \chi^2(\sin^2\vartheta_{13}),
\label{chi3nu}
\end{equation}
with the partial $\chi^2$'s extracted from Figure~3 of Ref.~\cite{Capozzi:2013csa}.
For each value of $m_{1}$
we calculated the confidence intervals for one degree of freedom.

Figure~\ref{3numbbvsmin} shows also the
90\% C.L. upper limit band for $|m_{\beta\beta}|$
estimated in Ref.~\cite{Bilenky:2014uka}
from the results of the
KamLAND-Zen experiment \cite{Gando:2012zm},
taking into account the uncertainties of the nuclear matrix element calculations.

In the left part of  Figure~\ref{3numbbvsmin} one can see that
there can be a complete cancellation of $|m_{\beta\beta}|$
for $m_{1}$ in the intervals that are given in Tab.~\ref{tab:3nu-nor}
at different confidence levels. Furthermore, $|m_{\beta\beta}|$
is larger than about 0.01 eV,
which is a value that
may be explored experimentally in the near future, for values of $m_{1} \gtrsim 0.008 \, \text{eV}$ which corresponds to almost degenerate
$m_{1}$ and $m_{2}$ (see Ref. \cite{Giunti:2015kza} for more details), hence, it will be very difficult to
measure $|m_{\beta\beta}|$
if there is a normal hierarchy of neutrino masses
($m_{1} \ll m_{2} \ll m_{3}$)
for any value of the unknown phases
$\alpha_2$
and
$\alpha_3$
in Eq.~(\ref{mbb3nu}).
We also note that 
$|m_{\beta\beta}| \gtrsim 0.01 \,\text{eV}$
is realized independently of
the values of the unknown phases
$\alpha_2$
and
$\alpha_3$
for
$m_{1} \gtrsim 0.04 \, \text{eV}$,
which is close to the region
$m_{1} \gtrsim 0.05 \, \text{eV}$
in which all the three neutrino masses are quasi degenerate (see Ref. \cite{Giunti:2015kza} for further details) .

In the case of Inverted Ordering ($m_{3} \ll m_{1} < m_{2}$)
there isn't any cancellation region of
$|m_{\beta\beta}|$
and we obtain
from the right part of Figure~\ref{3numbbvsmin}
the lower bounds
\begin{equation}
|m_{\beta\beta}|
>
1.6
\, (1\sigma)
,
1.5
\, (2\sigma)
,
1.3
\, (3\sigma)
\times 10^{-2} \, \text{eV}
.
\label{mbbminIO3nu}
\end{equation}
In the case of an Inverted Hierarchy ($m_{3} \ll m_{1} < m_{2}$) we also have the upper bounds
\begin{equation}
|m_{\beta\beta}|
<
4.8
\, (1\sigma)
,
4.9
\, (2\sigma)
,
4.9
\, (3\sigma)
\times 10^{-2} \, \text{eV}
.
\label{mbbmaxIO3nu}
\end{equation}
The next generations of
neutrinoless double-beta decay experiments
aim to explore the range of $|m_{\beta\beta}|$
between the limits found in Eqs.~(\ref{mbbminIO3nu}) and (\ref{mbbmaxIO3nu}) 
(see Refs.~\cite{GomezCadenas:2011it,Giuliani:2012zu,Schwingenheuer:2012zs,Cremonesi:2013vla,Artusa:2014wnl,Gomez-Cadenas:2015twa}),
and, in case the Inverted Hierarchy being the true one, in the next few years we will have a definitive answer of the Majorana nature of neutrinos.

\section{3+1 Mixing}
\label{sec:3p1}

The 3+1 mixing
is motivated
by the explanation of the
short-baseline anomalies \cite{Kopp:2013vaa,Giunti:2013aea},
which requires the existence of a new squared-mass difference
$\Delta{m}^2_{\text{SBL}} \sim 1 \, \text{eV}^2$.

In this section we will consider the case of 3+1 mixing
with a new massive neutrino $\nu_{4}$ at the eV scale, mainly sterile. In this case,
the effective Majorana mass in neutrinoless double-beta decay
is given by
\begin{equation}
|m_{\beta\beta}|
=
\left| m_{1}|U_{e1}| + m_{2}|U_{e2}| e^{i\alpha_2} + m_{3}|U_{e3}| e^{i\alpha_3}  + m_{4}|U_{e4}| e^{i\alpha_4} \right|,
\label{mbb3p1}
\end{equation}
and the contribution of $\nu_{4}$
enters with an unknown new phase
$\alpha_4$.

As in any extension of the standard $3\nu$ mixing, the ordering of the three lightest neutrinos are not known and then we will consider
separately the two cases
of Normal and Inverted Ordering of
$\nu_{1}$,
$\nu_{2}$,
$\nu_{3}$.


We calculated the confidence intervals using the $\chi^2$ function
\begin{equation}
\chi^2_{3+1}
=
\chi^2_{3\nu}
+
\chi^2(\Delta{m}^2_{\text{SBL}}, \sin^2\vartheta_{14})
,
\label{chi3p1}
\end{equation}
with
$\chi^2_{3\nu}$ defined in Eq.~(\ref{chi3nu})
and
$\chi^2(\Delta{m}^2_{\text{SBL}}, \sin^2\vartheta_{14})$
obtained from an update \cite{Giunti-NeuTel2015,GGLLLZ-15}
of the global fit of short-baseline neutrino oscillation
data presented in Ref.~\cite{Giunti:2013aea}.

\begin{figure}[t]
\begin{center}
\begin{minipage}{14pc}
\includegraphics[width=14pc]{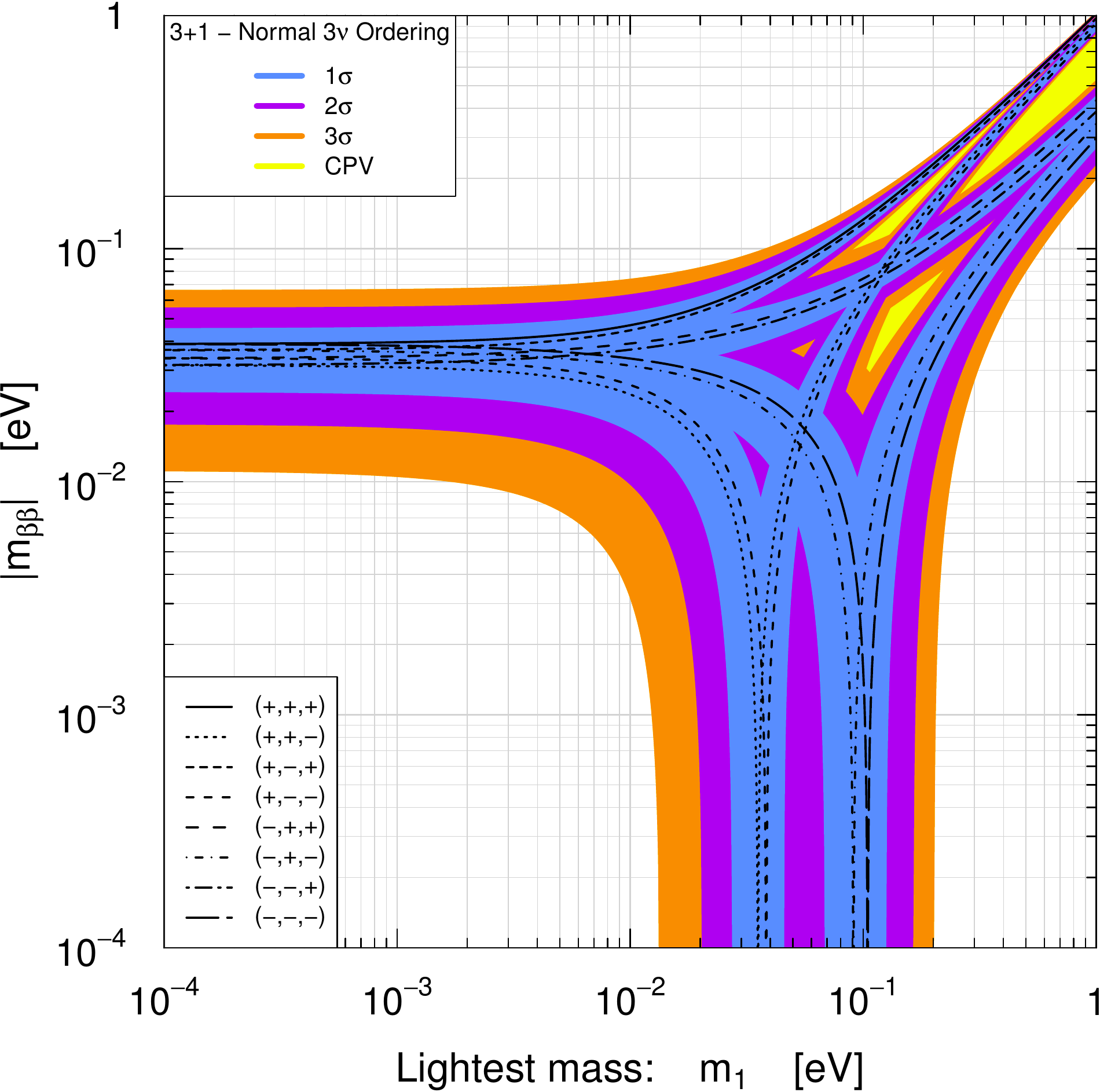}
\end{minipage}\hspace{2pc}%
\begin{minipage}{14pc}
\includegraphics[width=14pc]{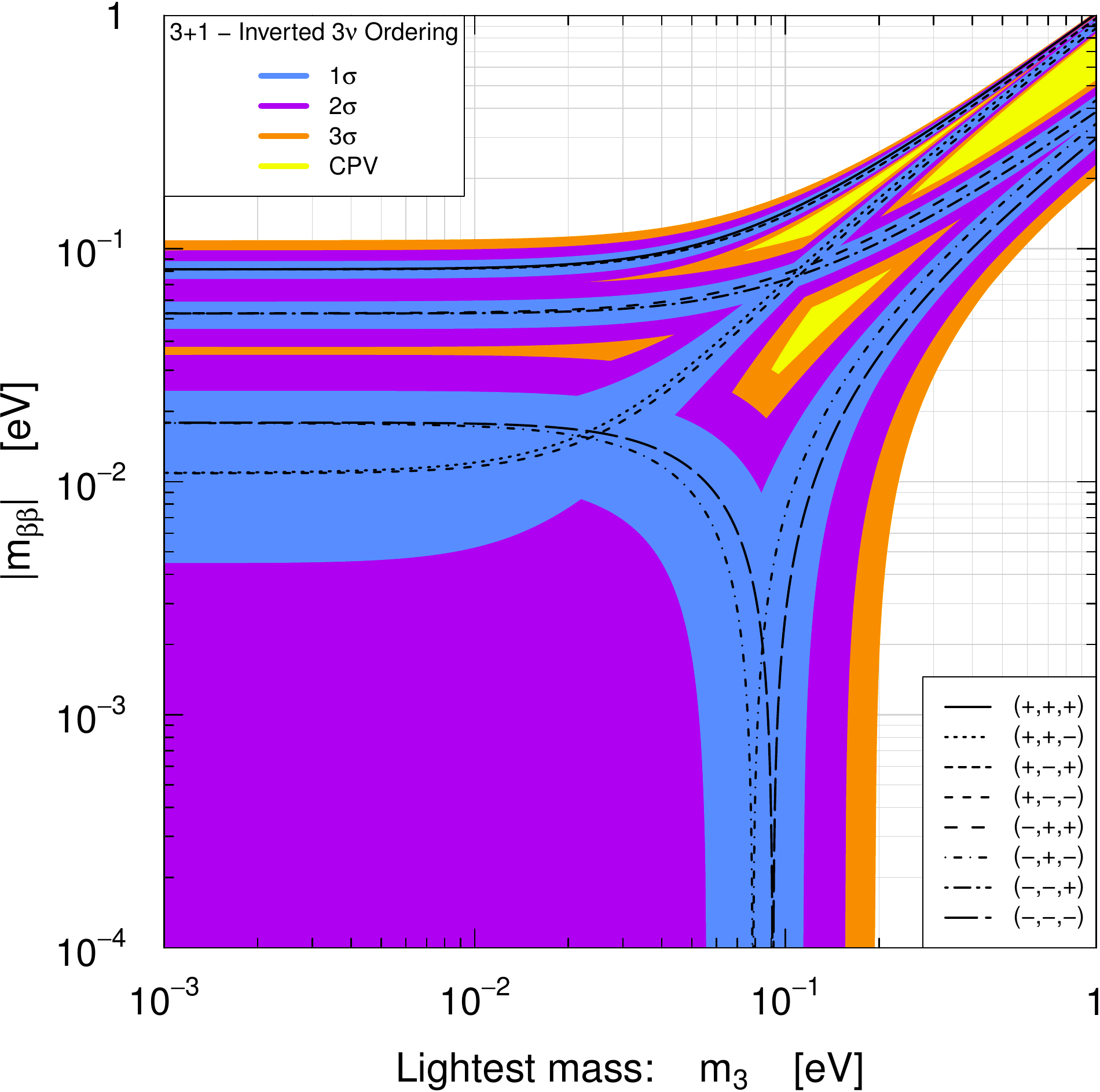}
\end{minipage}
\caption{Left part: Value of the effective Majorana mass $|m_{\beta\beta}|$
as a function of the lightest neutrino mass $m_{1}$
in the case of 3+1 mixing with Normal Ordering
of the three lightest neutrinos. Right part: Value of the effective Majorana mass $|m_{\beta\beta}|$
as a function of the lightest neutrino mass $m_{3}$
in the case of 3+1 mixing with Inverted Ordering
of the three lightest neutrinos.
The signs in the legend indicate the signs of
$e^{i\alpha_{2}}, e^{i\alpha_{3}} , e^{i\alpha_{4}} = \pm1$
for the four possible cases in which CP is conserved.
The intermediate yellow region is allowed only in the case of CP violation.}
\label{3p1mbbvsmin}
\end{center}
\end{figure}

\begin{figure}[t]
\begin{center}
\begin{minipage}{14pc}
\includegraphics[width=14pc]{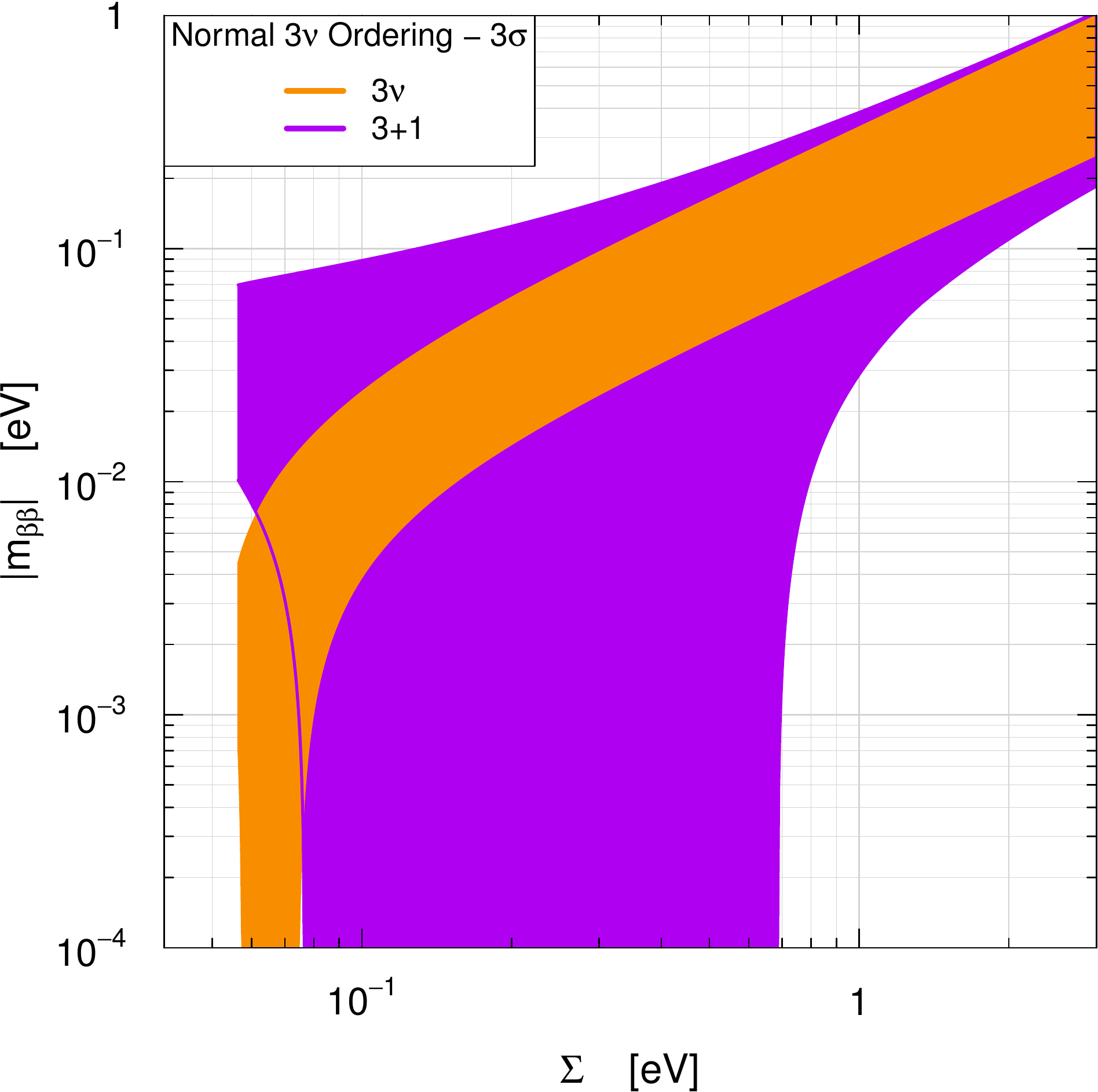}
\end{minipage}\hspace{2pc}%
\begin{minipage}{14pc}
\includegraphics[width=14pc]{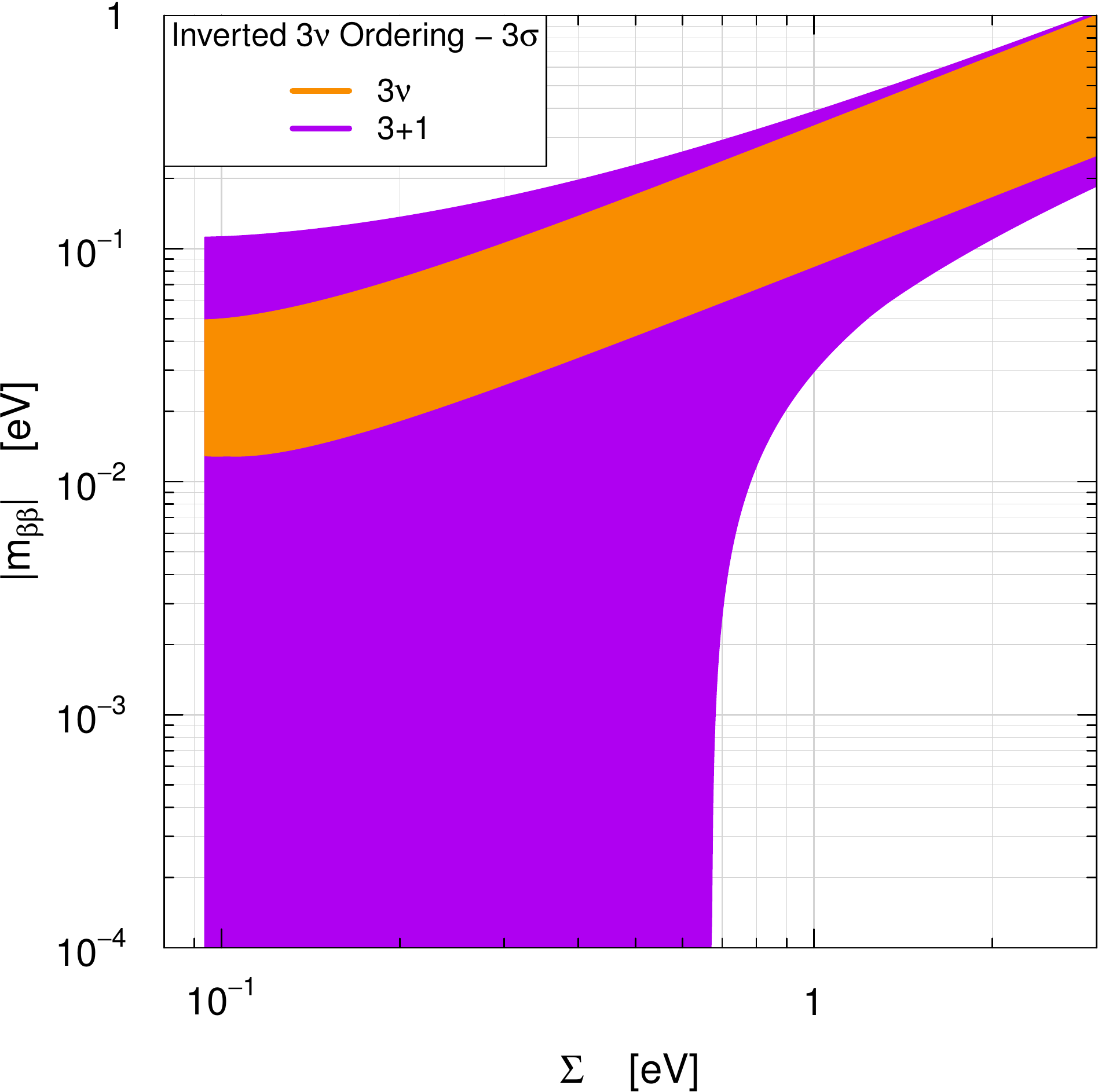}
\end{minipage}
\caption{Left part: Comparison of the $3\sigma$ allowed regions in the
$\Sigma$--$|m_{\beta\beta}|$ plane
in the cases of $3\nu$ and 3+1 mixing with Normal Ordering
of the three lightest neutrinos. Right part: Comparison of the $3\sigma$ allowed regions in the
$\Sigma$--$|m_{\beta\beta}|$ plane
in the cases of $3\nu$ and 3+1 mixing with Inverted Ordering
of the three lightest neutrinos.}
\label{compmbbvsmb}
\end{center}
\end{figure}

\begin{table}[b!]
\caption{Ranges of
$m_{1}$ \{$m_{3}$\}
for which there can be a complete cancellation of the
four partial mass contributions to $|m_{\beta\beta}|$
for the best-fit values (b.f.) of the oscillation parameters
and at
$1\sigma$,
$2\sigma$ and
$3\sigma$
in the case of 3+1 mixing with Normal Ordering \{Inverted Ordering\}
of the three lightest neutrinos.
}
\centering
\renewcommand{\arraystretch}{1.2}
\begin{tabular}{ccccc}
\hline
\hline
&
b.f.
&
$1\sigma$
&
$2\sigma$
&
$3\sigma$
\\
\hline
$m_{1}\,[10^{-2}\,\text{eV}]$
&
$3.5 - 10.5$
&
$2.8 - 12.5$
&
$2.0 - 16.3$
&
$1.3 - 20.0$
\\
\hline
$m_{3}\,[10^{-2}\,\text{eV}]$
&
$< 9.1$
&
$< 11.4$
&
$< 15.5$
&
$< 19.3$\\
\hline
\hline
\end{tabular}
\label{tab:3p1-nor}
\end{table}

The left part of Figure~\ref{3p1mbbvsmin} shows the allowed intervals for the best-fit value
and the
$1\sigma$,
$2\sigma$ and
$3\sigma$
of $|m_{\beta\beta}|$ for the 3+1 mixing with NO
as a function of $m_{1}$.
The corresponding intervals of
$m_{1}$ for which there can be a total cancellation of
$|m_{\beta\beta}|$
are given in Tab.~\ref{tab:3p1-nor}.

In Figure~\ref{3p1mbbvsmin}
we have plotted separately the
allowed bands for the eight possible cases in which CP is conserved
($\alpha_{2}, \alpha_{3}, \alpha_{4} = 0, \pi$).
The areas between the CP-conserving allowed bands
correspond to values of
$|m_{\beta\beta}|$ which are allowed only in the case of CP violation.

The right part of Figure~\ref{3p1mbbvsmin} presents the allowed regions in
the case for the 3+1 neutrino mixing  with IO as a function of $m_{3}$. Comparing Figure~\ref{3p1mbbvsmin}
with Figure~\ref{3numbbvsmin}
one can see that
the predictions for $|m_{\beta\beta}|$
are dramatically changed
from the $3\nu$ scheme to the 3+1 scheme
if there is an Inverted Ordering
of the three lightest neutrinos,
in agreement with the discussions in
Refs.~\cite{Goswami:2005ng,Goswami:2007kv,Barry:2011wb,Li:2011ss,Rodejohann:2012xd,Giunti:2012tn,Girardi:2013zra,Pascoli:2013fiz,Meroni:2014tba,Abada:2014nwa}.
The ranges of values of $m_{3}$
for which there can be a complete cancellation of $|m_{\beta\beta}|$
are given in Tab.~\ref{tab:3p1-nor}.

Figure~\ref{3numbbvsmin} gives a clear view of the possible values
of $|m_{\beta\beta}|$ depending on the scale of the lightest mass $m_{min}$
but, in practice,
the investigation of the absolute values of neutrino masses
is performed through the measurements
of the effective electron neutrino mass
in $\beta$-decay experiments \cite{Kraus:2004zw,Aseev:2011dq}
and through the measurement of the sum of the neutrino masses
\begin{equation}
\Sigma = m_{1} + m_2 + m_3
\label{sum}
\end{equation}
in cosmological experiments.
Hence it is interesting to compare the 3$\sigma$ allowed regions of $|m_{\beta\beta}|$ as a function of $\Sigma$ for the 3$\nu$ and 3+1 neutrino mixing schemes, 
as done in Figure~\ref{compmbbvsmb}.

The reason of this choice is that
$\Sigma$
is a measurable quantity also in the 3+1 scheme,
in cosmology,
the effects of the larger mass $m_{4}$
can be disentangled from those of the smaller masses,
because $\nu_{4}$ becomes non-relativistic
shortly after matter-radiation equality,
much earlier than
$\nu_{1}$,
$\nu_{2}$,
$\nu_{3}$.
In order to compute the confidence intervals we used the $\chi^2$ functions in Eq.~(\ref{chi3nu}) and  Eq.~(\ref{chi3p1})
with two degrees of freedom.


In the left part of Figure~\ref{compmbbvsmb} we show the comparison of the 3$\sigma$ allowed region in the $\Sigma$--$|m_{\beta\beta}|$ for the cases of 3$\nu$ and
3+1 mixing with Normal Ordering.
If the Normal Ordering will be established by oscillation experiments
(see Refs.~\cite{Bellini:2013wra,Wang:2015rma}),
with measurements of
$\Sigma$ and $|m_{\beta\beta}|$
it may be possible to distinguish $3\nu$ mixing and 3+1 mixing
if the measured values
select a region which is allowed only in one of the two cases.
It is interesting that there are two regions allowed only in the 3+1 mixing scheme:
one with $|m_{\beta\beta}|$ smaller than that in the case of $3\nu$ mixing
and
one with $|m_{\beta\beta}|$ larger than that in the case of $3\nu$ mixing.
Part of the second region can be probed in the next generation of neutrinoless double-beta decay experiments and it is a
very promising feature for distinguishing the 3+1 mixing and the $3\nu$ mixing schemes once the cosmological observations aim
to measure the sum of the three light neutrino masses down to the lower limit
of about
$5.6 \times 10^{-2} \, \text{eV}$
\cite{Audren:2012vy}.


The right part of Figure~\ref{compmbbvsmb}
shows the comparison of the $3\sigma$ allowed regions
in the $\Sigma$-$|m_{\beta\beta}|$ plane
in the cases of $3\nu$ and 3+1 mixing with Inverted Ordering
of the three lightest neutrinos.
If the Inverted Ordering will be established by oscillation experiments
(see Refs.~\cite{Bellini:2013wra,Wang:2015rma}),
it will be possible to exclude $3\nu$ mixing in favor of 3+1
by restricting
$\Sigma$ and $|m_{\beta\beta}|$
in the corresponding large region at small $|m_{\beta\beta}|$
allowed only in the 3+1 case.

\section{Conclusions}
\label{sec:conclusions}

We have presented some of the results obtained in Ref. \cite{Giunti:2015kza} in which we calculated the
effective Majorana mass $|m_{\beta\beta}|$
in neutrinoless double-$\beta$ decay
in the standard case of $3\nu$ mixing
and in the case of 3+1 neutrino mixing
indicated by the short-baseline anomalies.
We took into account the uncertainties of the
standard $3\nu$ mixing parameters
obtained in the global fit
of solar, atmospheric and long-baseline reactor and accelerator
neutrino oscillation data presented in
Ref.~\cite{Capozzi:2013csa}
and the uncertainties of the additional mixing parameters
in the 3+1 case
obtained from an update \cite{Giunti-NeuTel2015,GGLLLZ-15}
of the global fit of short-baseline neutrino oscillation
data presented in Ref.~\cite{Giunti:2013aea}.

We conclude that the predictions for $|m_{\beta\beta}|$
in the case of $3\nu$ is dramatically changed with the addiction of one sterile neutrino,
in agreement with the previous discussions in
Refs.~\cite{Goswami:2005ng,Goswami:2007kv,Barry:2011wb,Li:2011ss,Rodejohann:2012xd,Giunti:2012tn,Girardi:2013zra,Pascoli:2013fiz,Meroni:2014tba,Abada:2014nwa}.

We also compared the
allowed regions in the plane
$\Sigma$--$|m_{\beta\beta}|$,
taking into account the two possibilities of Normal and Inverted Ordering
of the three light lightest neutrinos.
We have shown that future measurements of these quantities
may distinguish the $3\nu$ and 3+1 cases
if the mass ordering is determined by oscillation experiments
(see Refs.~\cite{Bellini:2013wra,Wang:2015rma}).

\section{Acknowledgments}
E. Z. thanks the support of funding grants 2013/02518-7 and 2014/23980-3, S\~ao Paulo Research Foundation (FAPESP).
The work of C. Giunti is partially supported by the research grant {\sl Theoretical Astroparticle Physics} number 2012CPPYP7 under the program PRIN 2012 funded by the Ministero dell'Istruzione, Universit\`a e della Ricerca (MIUR).

\section*{References}

\bibliography{DBD}



\end{document}